\documentclass[conference,a4paper]{APSIPA2026}
\usepackage{amsmath}
\usepackage{graphicx}
\usepackage{multirow}
\usepackage{threeparttable}
\usepackage[backend=biber,style=ieee,]{biblatex}
\usepackage{amsmath}
\usepackage{amssymb}
\usepackage{booktabs}

\makeatletter
\let\labelindent\relax
\makeatother
\usepackage{enumitem}

\usepackage{geometry}
\usepackage{fancyhdr}

\fancypagestyle{firststyle}{
  \fancyhf{}
  \fancyhead[C]{2026 Asia Pacific Signal and Information Processing Association Annual Summit and Conference (APSIPA ASC)}
}

\newcommand{\Exp}{\mathbb{E}}

\usepackage{booktabs}
\usepackage{eso-pic}

\begin{document}

\AddToShipoutPictureFG*{%
  \AtPageLowerLeft{%
    \put(\LenToUnit{20mm},\LenToUnit{20mm}){%
      \parbox{170mm}{\centering\footnotesize
      \textcopyright{} 2026 IEEE. Personal use of this material is permitted.
      Permission from IEEE must be obtained for all other uses, in any
      current or future media, including reprinting/republishing this
      material for advertising or promotional purposes, creating new
      collective works, for resale or redistribution to servers or lists,
      or reuse of any copyrighted component of this work in other works.}
    }%
  }%
}

\title{KABURI-TTS: Phoneme-Keyed Activity-conditioned Bi-channel Utterance Rendering for Interaction}

\author{
\authorblockN{
Ryuichiro Higashinaka\authorrefmark{1},\authorrefmark{2},
Shinnosuke Takamichi\authorrefmark{3},
Tetsuji Ogawa\authorrefmark{4}
}
    
\authorblockA{\authorrefmark{1} Nagoya University, Japan. higashinaka@i.nagoya-u.ac.jp}
\authorblockA{\authorrefmark{2} National Institute of Informatics, LLMC, Japan.}
\authorblockA{\authorrefmark{3} Keio University, Japan.}
\authorblockA{\authorrefmark{4} Waseda University, Japan.}
}

\maketitle
\thispagestyle{firststyle}
\pagestyle{empty}

\begin{abstract}
Realizing full-duplex spoken dialogue requires large amounts of two-channel, one-speaker-per-channel conversational speech data. Although conversational text-to-speech (TTS) engines have been developed, they are not necessarily robust to two-party simultaneous phenomena such as backchannels, interruptions, and overlaps that occur while the interlocutor is speaking. In this work, aiming at conversational speech synthesis that reproduces human-like overlap, we propose KABURI-TTS. KABURI-TTS takes a per-speaker phoneme raster as input and renders the speech of the two speakers on separate channels, conditioned on the per-frame phonemes and the voice activity derived from them. Because the phoneme raster is supplied by a separate module, the proposed method enables controllable generation of one-speaker-per-channel, two-party spoken dialogue. A user evaluation shows that, compared with strong baselines, the proposed method attains higher naturalness at both the utterance and the interaction level. Furthermore, an analysis of voice activity confirms that the proposed method produces more overlap and more frequent turn-taking.
\end{abstract}


\section{Introduction}
\label{sec:introduction}

In recent years, spoken dialogue models have advanced to the point where full-duplex spoken dialogue can be realized \cite{nguyen-etal-2023-generative,defossez2024moshi}. While such models achieve natural, human-like interaction, their training requires large amounts of spoken dialogue data. In particular, a large amount of two-channel, one-speaker-per-channel dialogue data is needed, and various techniques have been employed to obtain it, such as applying speaker separation to monaural recordings. Speech synthesis is also widely used for this purpose. Because synthesis allows a system to be adapted to a desired domain, it has been adopted in many systems \cite{lee-etal-2025-behavior,roy2026personaplexvoicerolecontrol}.

However, such two-channel speech synthesis still faces many challenges. The most significant problem is that these methods are not necessarily robust to two-party simultaneous speech, such as backchannels, interruptions, and overlaps that occur while the interlocutor is speaking. Existing approaches include synthesizing speech separately for each speaker and then arranging the utterances so that the result sounds natural \cite{lee-etal-2025-behavior}, modeling the speech of both speakers from text input \cite{borsos2023soundstormefficientparallelaudio,zhang2026mossttsdtextspokendialogue}, and generating dialogue speech by leveraging an existing full-duplex spoken dialogue model \cite{defossez2024moshi}. However, these methods can produce unnatural dialogue, and because overlap is not explicitly conditioned on, they sometimes generate unnatural dialogue speech with little overlap. Moreover, given the goal of training spoken dialogue models, it is desirable to be able to control how simultaneous speech is produced; however, existing methods learn this implicitly, which makes such control difficult.

In this work, in light of these problems, we propose KABURI-TTS: phoneme-{\bf K}eyed {\bf A}ctivity-conditioned {\bf B}i-channel {\bf U}tterance {\bf R}endering for {\bf I}nteraction. {\it Kaburi} is a Japanese word meaning simultaneous speech. KABURI-TTS takes a per-speaker phoneme raster as input and generates the speech of the two speakers on separate channels, conditioned on the per-frame phonemes and their voice activity. The model is built on top of an existing single-channel, single-speaker speech synthesis engine, and is trained on human-to-human spoken dialogue data annotated with phoneme rasters. By designing the phoneme raster to be supplied from a separate module, the phonemes and the duration of utterances can be customized. This makes it possible to generate controllable, one-speaker-per-channel, two-party spoken dialogue that reliably produces simultaneous speech.
We trained a Japanese speech synthesis model on Japanese casual conversation dialogue data. A human evaluation confirmed that, compared with competitive baselines, the proposed method improves the naturalness of dialogue. We also confirmed that simultaneous speech can be customized by changing how the phoneme raster is constructed. 
The code and pretrained models are available at {\tt https://github.com/llm-jp/kaburi-tts}.

\section{Related Work}
\label{sec:related_work}

Conversational speech synthesis is currently attracting considerable attention. These methods can be broadly divided into three lines of work. 

The first uses a single-speaker TTS to generate the speech of each speaker utterance by utterance, and then constructs the dialogue speech by placing these utterances \cite{lee-etal-2025-behavior}. Because single-speaker TTS produces high-quality speech, the audio quality is very high; however, this approach has the limitation that it cannot generate naturally interactive speech or handle the interaction between speakers. 

The second serializes speaker-tagged utterances and produces a single mixed-speech signal from them. Starting with SoundStorm \cite{borsos2023soundstormefficientparallelaudio}, methods such as MOSS-TTSD \cite{zhang2026mossttsdtextspokendialogue} have been proposed. While this makes it easier to generate natural conversational speech, the serialized input means that simultaneous speech is not learned explicitly, and the output tends to  alternate. Moreover, simultaneous speech is difficult to control. CoVoMix \cite{NEURIPS2024_b5fd95d6}, which handles per-speaker semantic token sequences, can model the interaction between speakers, but it cannot control simultaneous speech and generates speaker-mixed audio.

The third generates the speech of the two speakers on separate channels from text input. For example, CHATS \cite{mitsui2023humanlikespokendialoguegeneration}, Multistream TTS based on Moshi \cite{defossez2024moshi}, ZipVoice-Dialogue-Stereo \cite{zhu-etal-2026-zipvoice}, and DialoSpeech \cite{xie2025dialospeechdualspeakerdialoguegeneration} can generate one-speaker-per-channel speech for two speakers from text. However, because these methods also generate speech directly from text, which speech overlaps and how it overlaps are learned implicitly from data, and the problems remain that simultaneous speech does not appear sufficiently or cannot be controlled. 

This work aims to realize a TTS for spoken dialogue that can reliably represent simultaneous speech based on an explicit phoneme raster, thereby addressing the existing problems of insufficient and hard-to-control simultaneous speech. Moreover, by designing the phoneme raster to be supplied externally, we make it easier for the designer to control overlap.

\section{Method}
\label{sec:method}

We describe the problem setting, the model, and the training method of KABURI-TTS. In addition, since a phoneme raster is required to actually use the model as a TTS, we describe statistical placement and a duration predictor as mechanisms for producing it. In this work, we use DACVAE \cite{niu2026semanticvaesemanticalignmentlatentrepresentation}, a neural audio codec that encodes speech into a sequence of $d$-dimensional continuous latents, one per frame, and decodes them back into a waveform. As the backbone, we use a pre-trained single-speaker zero-shot TTS (a 12-layer Rectified-Flow DiT), and we adapt it to two-speaker stereo dialogue by using LoRA and updating some of the weights.

\subsection{Formulation}

The inputs are the phoneme and activity rasters and the speaker references.
We discretize a dialogue into $T$ frames at $25\,\mathrm{fps}$, and let
$t\in\{1,\dots,T\}$ be the frame index and $c\in\{A,B\}$ be the channel
representing the two speakers. All inputs are sequences synchronized with
these $T$ frames, which we call rasters.
For speakers $A$ and $B$, we define the phoneme raster $\mathbf p=(p^A,p^B)$,
where $p^c_t\in\mathcal V\cup\{\textsf{sil}\}$ is the phoneme uttered by channel
$c$ at frame $t$. Here, $\mathcal V$ is the phoneme vocabulary and $\textsf{sil}$
is the silence token. We define the activity raster $\mathbf a=(a^A,a^B)$,
where $a^c_t\in\{0,1\}$ represents who speaks when, and $a^c_t=1$ indicates
that channel $c$ is speaking at frame $t$. The speaker reference is a latent
$s=(s^A,s^B)$, where $s^c\in\mathbb{R}^{T_\mathrm{ref}\times d}$ is the latent
obtained by encoding the reference speech of channel $c$ with DACVAE. 

The output is the per-frame, per-channel latent: each speaker channel is a
DACVAE latent sequence $z^c\in\mathbb{R}^{T\times d}$ ($d{=}32$), the whole
dialogue is $z=(z^A,z^B)$, and each channel is decoded into a waveform
$x^c=\mathrm{Dec}(z^c)$ by the DACVAE decoder.

The acoustic model defines the conditional distribution of the
stereo latents, $p_\theta\big(z\mid\mathbf p,\mathbf a,s\big)$.

\subsection{Model}

\begin{figure}[t]
\begin{center}
\includegraphics[width=0.44\textwidth]{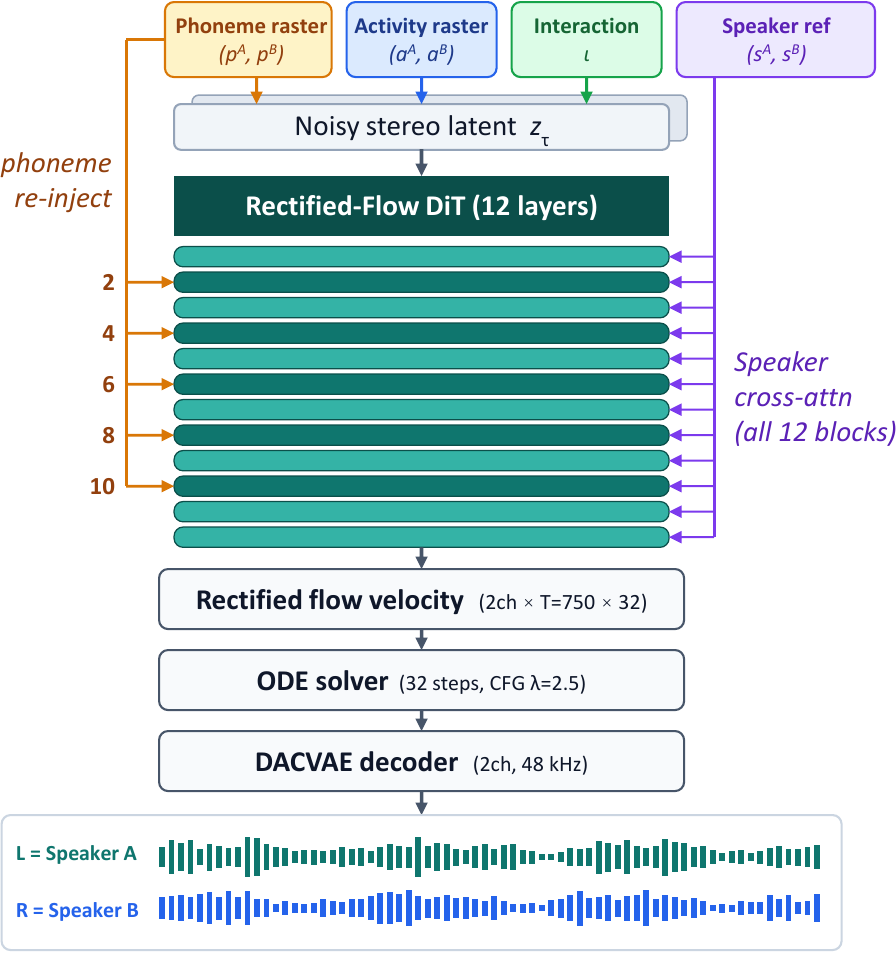}
\end{center}
\caption{\label{fig:arch}Model architecture}
\vspace*{-5pt}
\end{figure}

The architecture of the acoustic model is shown in Figure~\ref{fig:arch}.
There are four types of inputs: the phoneme raster $(p^A,p^B)$, the activity
raster $(a^A,a^B)$, the interaction state $\iota$, and the speaker reference
$(s^A,s^B)$. Although the activity raster can in general be provided
independently, in this work we obtain it from the phoneme raster as
$a = \mathbf{1}[p \neq \mathsf{sil}]$. The interaction state $\iota$ is a
function $\iota=g(\mathbf a)$ computed deterministically from the activity
raster $\mathbf a=(a^A,a^B)$ (speech/silence at each frame). We classify each
frame into a base state determined solely by the current frame (four types:
silence, A only, B only, and overlap) and a boundary state determined by the
difference from the previous frame (six types in total: the onset and offset
of each speaker's speech, and speaker change A$\to$B / B$\to$A), and assign a
total of ten classes $\iota_t\in\{0,\dots,9\}$. At frames where a boundary
occurs, the boundary state takes precedence over the base state. In this way,
we extend the activity raster and explicitly condition on the voice activity
state.

The generation target is the noisy version $z_\tau$ of the stereo latent 
$z=(z^A,z^B)$, which the backbone Rectified-Flow DiT (12 layers) maps to a
velocity field. The phoneme, activity, and interaction conditions are injected as
additive biases after projecting the latents to the hidden space. In addition,
the phoneme condition is re-injected into the deep blocks (layers 2, 4, 6, 8, and
10) so that the phoneme information is retained throughout the layers. Speaker
characteristics are reflected by feeding the reference latents to the speaker
cross-attention of all twelve blocks. The two channels $(A,B)$ are stacked
along the batch dimension and processed simultaneously in a single forward pass
with shared weights. The two channels are not directly mixed; their interplay
is represented through the conditions described above. 
In this work, we target 30-second chunks, so the number of frames is
$T{=}750$ at $25\,\mathrm{fps}$. 

The model output is the rectified-flow velocity $v_\theta$ (of shape
$2\text{ch}\times T{=}750\times 32$). At inference time, we obtain the latent
$\hat z$ by 32-step ODE numerical integration (classifier-free guidance,
$\lambda{=}2.5$), and decode it into a waveform with the DACVAE decoder (2ch,
48\,kHz). The left channel of the output stereo corresponds to speaker A, and
the right channel to speaker B.

\subsection{Training}

We adapt the backbone by inserting low-rank
adaptation (LoRA, $r=256$) into the attention and MLP layers, and update only
the later blocks, the LayerNorms, and the input/output layers by unfreezing
them. The newly added condition-injection layers are zero-initialized so that,
before training, the output remains equivalent to that of the backbone. 

The velocity field $v_\theta$ is trained using rectified flow
\cite{liu2022flowstraightfastlearning}. The total training loss consists of a
velocity regression term as the main term, together with terms that suppress
deviation in timing and deviation from the base model.
\begin{equation}
\begin{aligned}
\mathcal L=\;
& \Exp_{z,\epsilon,\tau}\!\big[\,\omega(\mathbf a)\,\lVert v_\theta(z_\tau,\tau,p,\mathbf a,s)-(\epsilon-z)\rVert^2\big]\\
& {}+\textstyle\sum_k \lambda_k\,\mathrm{BCE}(h_k,y_k)
  +\lambda_{\mathrm{sp}}\lVert\theta_{\mathrm{ft}}-\theta_0\rVert^2 .
\end{aligned}
\end{equation}
Here, $\tau\in[0,1]$ is the rectified-flow time and
$z_\tau=(1-\tau)z+\tau\epsilon$.
The first term is the main loss, namely velocity regression on the latents.
Here, to suppress the contribution of silent frames, we introduce a per-frame
weight $\omega(\mathbf a)=\mathbf a+0.05\,(1-\mathbf a)$ and focus the training
on the speech portions. Here, $\mathbf a=(a^A,a^B)$ is the activity raster, and
$a^c_t\in\{0,1\}$ indicates whether the speaker on channel $c$ is speaking at frame $t$ (1) or
not (0). The reason we do not set silence completely to $0$ is that the latents
of silence also need to be generated.

The second term is an auxiliary loss for timing. From the intermediate
representations of the DiT, small heads predict the timing elements of the
dialogue, each in a binary manner. There are five prediction targets: activity
is whether each speaker is speaking, onset/offset are the start and end frames
of each speaker's speech, overlap is whether the two speakers are speaking
simultaneously, and turn-switch is whether a speaker change has occurred
(without distinguishing its direction); activity and onset/offset are judged for
each of the channels of speakers A and B. We match each prediction $h_k$ against
the binary label $y_k$ of the same kind derived from the ground-truth activity
raster $\mathbf a$ using binary cross-entropy to obtain the auxiliary loss. The
weight of each auxiliary loss is $\lambda_k = 0.1$ (equal for all five types).

The third term is L2-SP regularization. It pulls the unfrozen weights to be
updated $\theta_{\mathrm{ft}}$ back toward their pre-trained values $\theta_0$
($\lambda_{\mathrm{sp}}=10^{-4}$), preventing the forgetting of the knowledge
held by the base model. 

Note that, for classifier-free guidance, we drop the phoneme condition with probability $0.15$ during training, and apply guidance (scale $=2.5$) to it at inference.

\subsection{Construction of the Phoneme Raster}
Since the proposed model takes a phoneme raster as input, it cannot be used
directly as a TTS that takes text as input. We therefore need to additionally
employ a method for constructing a phoneme raster from text. Here, we prepare two
such methods: a method that performs simple statistical placement, and a method
that trains a duration predictor.

\subsubsection{Statistical Placement}

Statistical placement assigns a phoneme raster to text using only training-data statistics without learning.
In this work,
inspired by prior work \cite{lee-etal-2025-behavior}, we place each utterance in
time on its corresponding channel, and determine two quantities as follows: the
gap between utterances and the phoneme durations within an utterance.

The start time of an utterance is determined by the gap from the end of the
preceding utterance. We provide this gap by inverse-CDF sampling from the
direction-dependent empirical distributions (A$\to$A, A$\to$B, B$\to$A,
B$\to$B) of the training data. Specifically, we construct a piecewise-linear
inverse CDF whose knots are the quantiles
$(p_{05},p_{10},\mathrm{med},p_{90},p_{95})$ of the gap distribution for each
direction, and obtain a gap value through a uniform random variable
$u\sim\mathcal U(0.05,0.95)$. By drawing directly from the empirical
distributions, we can reproduce the turn-taking distribution (gaps and overlaps)
of real dialogue without assuming, for example, a Gaussian distribution or using
hand-set constants. We restrict the uniform random variable to $0.05$--$0.95$ in
order to exclude extreme outliers (excessively long silences or overlaps). In
speaker changes, a negative gap, that is, an overlap, arises naturally; to avoid
excessive overlap, we set an upper bound of $0.3$\,s.

As for the utterance length, we first determine the number of frames of the
whole utterance as $D=n\cdot\widetilde{\mathrm{fpp}}$, where $n$ is the number
of phonemes and $\widetilde{\mathrm{fpp}}$ is the median number of frames per phoneme
in the training data, and we then distribute it to each phoneme $i$ in proportion
to the training-average phoneme duration $\bar d_{\pi_i}$:
$d_i = D\,\bar d_{\pi_i}/\sum_k \bar d_{\pi_k}$. If the resulting sequence of
times exceeds the window length $T$, we uniformly scale the whole to fit. From
the above, we construct the phoneme raster of the two channels and feed it to the
acoustic model.

\subsubsection{Duration Predictor}
For more flexible timing generation, we also prepare a lightweight predictor
that predicts timing directly from text. This is a small Transformer encoder (2
layers, hidden dimension 192, about $0.8$\,M parameters) that takes as input
token sequences of phonemes, speaker, channel, and dialogue context, and predicts
each phoneme duration, each silence duration, and the gap between utterances as a
classification problem. The numbers of bins are 30 for phoneme duration, 24 for
silence duration, and 57 for gap (where negative values represent overlap).

For training, in addition to the cross-entropy against the ground truth, we use
auxiliary losses on utterance length, speaking rate, and frame state. At
inference, we convert the predicted bins back into continuous values by the
expectation (for phoneme duration and gap) or the argmax (for silence duration),
and assemble them into the phoneme and activity rasters of the two channels. If the
result exceeds the window length, we scale it to fit. Note that the predicted
gaps tend to shrink toward the center due to the expectation decoding, which
makes turn-taking too tight; we therefore add a uniform correction of $+9$
frames, calibrated on the validation data, to the gap between utterances.

\section{Experiment}
\label{sec:experiment}
To verify the effectiveness of KABURI-TTS, we conducted a human evaluation. We
performed two experiments. The first is speech synthesis on test chunks
({\bf Test chunk Eval}). The second, as a more practical evaluation, is speech
synthesis on separately created spoken dialogue text ({\bf Text-in Eval}). We
prepared competitive baselines and compared the proposed method against them.
For the proposed method, we used Irodori
TTS\footnote{\url{https://github.com/Aratako/Irodori-TTS}}\footnote{\url{https://huggingface.co/Aratako/Irodori-TTS-500M-v2}}
as the backbone model of the DiT. We disabled the text input used by Irodori TTS
by setting its weights to zero.

\subsection{Dataset}
For the training data of KABURI-TTS, we used LLM-jp-Zoom1, a large-scale
Japanese casual conversation spoken dialogue dataset \cite{zoom1}. This
dataset contains 2{,}000 approximately 30-minute speech conversations recorded
over Zoom, amounting to 1{,}085 hours. The number of participating speakers is
60. The speakers chat about 15 topics, such as beauty and sports. We split this
data into 30-second chunks and then created train/valid/test splits. 

For the
test data, in order to be able to verify the handling of unseen speakers, we
created subsets that include both seen and unseen speakers. However, because
there are only 60 speakers in total, it is difficult to create a split
consisting entirely of mutually unseen speakers. Therefore, the test-unseen
subset contains 10 completely unseen speakers and 7 seen speakers. 

We set valid, test-seen, and test-unseen to 80 dialogues each, and used the
remainder as the train split. The train split contains 1{,}760 dialogues
(113{,}310 chunks, 50 speakers), and the valid, test-seen, and test-unseen
splits contain 5{,}159, 5{,}131, and 5{,}128 chunks, with 43, 39, and 17
speakers, respectively.

All dialogue speech was segmented into utterances with Silero
VAD\footnote{\url{https://github.com/snakers4/silero-vad}} and transcribed with
Whisper-v3\footnote{\url{https://github.com/openai/whisper}}. Then, for each
chunk, we applied g2p to the transcription results and performed alignment using
the Montreal Forced Aligner \cite{mcauliffe17_interspeech} to obtain a phoneme
raster for each channel. When there was speech at a chunk boundary, the
transcription result on that boundary was included in the chunk.

In Test chunk Eval,
we sample test chunks for evaluation. Specifically, we randomly selected 50
samples from test-seen and 50 samples from test-unseen, preparing 100 test
chunks in total. Using these phoneme rasters as input, we synthesize dialogue
speech with the proposed model and evaluate it.

In Text-in Eval, we had an LLM generate dialogue text according to the following
instructions. The LLM was instructed to write a free two-speaker dialogue in
Japanese on a given topic, as everyday casual conversation of 16 utterances,
satisfying the following constraints:
\begin{itemize}
    \item The speakers are A and B, and each line contains one utterance in the
    form ``A: text'' / ``B: text''.
    \item The utterances are not strictly alternating; places where the same
    speaker continues twice are inserted naturally.
    \item About three to five short backchannels are included.
    \item Hard-to-read items such as proper nouns, alphabetic characters, and
    numbers are not used.
    \item The utterances are short and in spoken language rather than written
    language (each utterance is assumed to be roughly 5 to 25 characters).
    \item The content is a coherent casual conversation.
\end{itemize}
These guidelines were obtained by observing the train chunks. Using OpenAI
GPT-5.5 with the high reasoning setting, we generated four text dialogues for
each of the 15 topics in LLM-jp-Zoom1, creating 60 inputs in total.

\subsection{Systems for Comparison}
For the speech synthesis on test chunks, we prepared the following systems for
comparison. For KABURI-TTS, the acoustic model was trained on 7 Quadro RTX 6000
GPUs with an effective batch size of 28 for 30k steps (about 7.4 epochs), and we
used the final checkpoint. The optimizer was AdamW (learning rate
$1\times10^{-4}$). The duration predictor was trained on 1 Quadro RTX 6000 GPU
with a batch size of 32 for 12k steps (about 3.4 epochs), and we used the final
checkpoint. The optimizer was AdamW (learning rate $5\times10^{-5}$). For all
systems, the same speaker reference information is used.
\begin{description}[style=sameline, leftmargin=1em, font=\bfseries]
    \item[GT]{As an upper bound, this uses the original human-to-human speech
    as it is.}
    \item[KABURI (GT)]{The proposed method, which generates speech using
    the GT phoneme raster of the test chunk directly as input.}
    \item[KABURI (stat)]{A method that takes the collapsed phoneme sequence
    obtained by applying g2p to the text of the test chunk as input, constructs a
    phoneme raster by statistical placement, and generates speech with the acoustic
    model.}
    \item[KABURI (pred)]{The same as above, but the phoneme raster is
    constructed by the trained duration predictor instead of statistical
    placement.}
    \item[Irodori (GT)]{Speech is generated by feeding each utterance
    text into Irodori TTS, and the start positions of the utterances are set to
    the GT placement. This serves as a baseline.}
    \item[Irodori (stat)]{The same as above, but the start positions of
    the utterances are set by statistical placement.}
\end{description}
For Text-in Eval, in addition to KABURI (stat), KABURI (pred), and Irodori
(stat), we prepared MOSS-TTSD \cite{zhang2026mossttsdtextspokendialogue}.
MOSS-TTSD is a model that is regarded as high-performing for generating dialogue
speech from text, and can be considered a competitive baseline. We fed the
evaluation text and reference speech into it to generate speech. Note that, since
MOSS-TTSD is a model that generates single-channel speech, its evaluation target
is, strictly speaking, different from those of the other methods.
For reference speech, each Text-in Eval dialogue used one speaker pair sampled from the top-10 training speakers in LLM-jp-Zoom1 for all systems.

\subsection{Evaluation Metrics}

The human subjective evaluation used three items, each rated on a five-point
Likert scale (1 being the worst and 5 being the best). Q1 (utterance-level
naturalness) asks whether each utterance has a natural speaking style, like that
heard in everyday conversation between humans. Q2 (interaction-level naturalness)
asks whether the connection between utterances sounds natural, as the exchange
of a natural conversation. Q3 (intelligibility) asks whether the content of the
conversation can be heard without difficulty.

\subsection{Evaluation Procedure}
For the evaluation, we used CrowdWorks, a crowdsourcing platform. For each test
sample (the speech generated by each method for the 100+60=160 inputs, each
about 25 to 40 seconds depending on the system), two crowd workers were
assigned, and they evaluated it on the Likert scales described above. A total of
140 different workers participated, and each worker evaluated 12 speech samples.
The evaluation experiment was approved by the ethics review board of our
institution.

\subsection{Results}

Table~\ref{tab:evalA} shows the results of Test chunk Eval. The comparison
between methods was performed by the Wilcoxon signed-rank test (paired) on the
item-level scores, which were obtained by averaging the ratings of the two
subjects for each item. For multiple comparisons, we treated each evaluation
item as a family, and within each family we applied the Benjamini--Hochberg
procedure to control the FDR at 5\%.

First, it is clear that GT is high overall and that the synthesized speech does
not reach it. In addition, KABURI performs well when the GT placement is used.
This confirms that appropriate speech can be generated when the ground-truth
phoneme raster is given as input. On top of this, regarding the naturalness of
utterances, among the synthesized speech, KABURI (GT) is good, but KABURI (pred)
is also good. It is significantly better than KABURI (stat) and the
Irodori-based methods. 
One possible reason for the higher utterance-level naturalness of KABURI (pred) is that statistical placement assigns phoneme durations based on corpus-level statistics, whereas the duration predictor uses phoneme, speaker, and dialogue-context information. This may allow KABURI (pred) to better model context-dependent within-utterance timing in Test chunk Eval.
Since there is a significant difference from KABURI (GT),
the placement by the predictor is not sufficient, but we can say that it achieves
a sufficiently natural placement.
Regarding the naturalness of the interaction, although the GT placement is good,
KABURI (pred) and KABURI (stat) are comparable and better than the Irodori-based
methods. This shows that the proposed method can sufficiently achieve the
naturalness of the interaction. Moreover, in this respect, we find that
statistical placement is sufficiently strong.
Regarding intelligibility, KABURI (stat) is better than KABURI (pred) and the
Irodori-based methods, and is numerically better than KABURI (GT).

\begin{table}[t]\centering
\caption{Evaluation results in Test chunk Eval. Among the methods excluding GT,
\textbf{bold} indicates the best value in each column, and \underline{underline}
indicates the second best. Each row is labeled a--f, and the superscript symbols
to the right of a value indicate which methods that method is significantly
better than. Two identical letters (e.g., $^{cc}$) denote $p<.01$, and a single
letter (e.g., $^{c}$) denotes $p<.05$.}
\label{tab:evalA}
\setlength{\tabcolsep}{1mm}
\begin{tabular}{lllll}
\toprule
 & System & Utt. naturalness & Inter. naturalness & Intelligibility \\
\midrule
a & GT (upper bound) & 4.42$^{bbccddeeff}$ & 4.43$^{bbccddeeff}$ & 4.31$^{bbccddeeff}$ \\
b & KABURI (GT)   & \textbf{3.49}$^{ccddeeff}$ & \textbf{3.64}$^{ddeeff}$ & \underline{3.65} \\
c & KABURI (stat) & 2.96 & \underline{3.46}$^{eeff}$ & \textbf{3.82}$^{ddeeff}$ \\
d & KABURI (pred) & \underline{3.20}$^{cef}$ & 3.37$^{eeff}$ & 3.54 \\
e & Irodori (GT)   & 2.90 & 2.61 & 3.49 \\
f & Irodori (stat) & 2.89 & 2.79$^{e}$ & 3.54 \\
\bottomrule
\end{tabular}
\vspace*{-5pt}
\end{table}

We also compared the test-seen and test-unseen subsets for each system and axis (Mann–Whitney U test with BH–FDR correction) and found no significant differences.

Table~\ref{tab:evalB} shows the evaluation results of Text-in Eval. First,
throughout the results, we can see that KABURI (stat) achieves good scores.
MOSS-TTSD is the second best in the naturalness of the interaction and in
intelligibility, and is relatively good. Regarding the naturalness of utterances,
although no significant difference is observed, 
we can see that KABURI obtains higher scores than the baselines.
Regarding the naturalness of the interaction, KABURI
(stat) is significantly better than KABURI (pred) and Irodori (stat). Although
MOSS-TTSD shows no significant difference from any of the KABURI variants, for
the naturalness of the interaction, a marginally significant tendency was
observed in favor of KABURI (stat) at a corrected $p=0.071$. 
From this, the proposed method can be considered more effective than the baselines overall.
Note that MOSS-TTSD generates mixed speech, unlike the proposed method. Despite this more difficult setting, the proposed method achieves comparable or superior interaction naturalness.

\begin{table}[t]\centering
\caption{Evaluation results of Text-in Eval. \textbf{Bold} indicates the best
value in each column, and \underline{underline} indicates the second best. Each
row is labeled a--d, and the superscript symbols to the right of a value indicate
which methods that method is significantly better than. For the notation, refer
to the previous table.}
\label{tab:evalB}
\setlength{\tabcolsep}{1.2mm}
\begin{tabular}{lllll}
\toprule
 & System & Utt. naturalness & Inter. naturalness & Intelligibility \\
\midrule
a & KABURI (stat)  & \textbf{3.38} & \textbf{3.83}$^{bbcc}$ & \textbf{3.86} \\
b & KABURI (pred)  & \underline{3.30} & 3.30 & 3.76 \\
c & Irodori (stat)  & 3.07 & 3.11 & 3.72 \\
d & MOSS-TTSD & 3.20 & \underline{3.53}$^{c}$ & \underline{3.79} \\
\bottomrule
\end{tabular}
\vspace*{-5pt}
\end{table}

For practical text-to-speech use, the Text-in Eval results matter most, where KABURI (stat) is the best overall. 
KABURI (pred) attains high utterance naturalness in Test chunk Eval, suggesting room for improvement with a better predictor.

\section{Analysis}
\label{sec:analysis}

We additionally analyzed how much simultaneous speech KABURI-TTS can actually
achieve. For each channel, we computed the short-time energy in a 20 ms window
at 25 fps and regarded frames with at least 5\% of the channel's peak energy as
speech, obtaining a binary voice activity sequence. From this, we measured
Overlap (the proportion of frames in which both channels speak simultaneously),
Silence (the proportion in which neither speaks), and Switches/min (the
per-minute number of changes of the solely speaking speaker). Larger Overlap and
Switches/min indicate denser interplay and more frequent exchanges, while larger
Silence indicates longer pauses. Table~\ref{tab:turntaking} shows the results.

In Test chunk Eval, KABURI (GT) achieves turn-taking close to GT, and KABURI
(stat) and KABURI (pred) are also close to GT, although KABURI (pred) takes more
pauses than KABURI (stat). The Irodori-based methods have little overlap, much silence, and few turns: because each utterance is synthesized independently, utterance lengths do
not match the real speech, and short responses or backchannels are absorbed into
the interlocutor's speech or concatenated with adjacent same-speaker speech, so
short turn changes rarely appear as a single floor. In Text-in Eval, KABURI
attains much overlap and frequent turn-taking, whereas the Irodori-based methods
produce non-overlapping, block-like utterances. 
These metrics support the higher subjective naturalness of KABURI.
Note that these statistics characterize aggregate output activity and do not directly measure frame-level adherence to the input raster. Frame-level control fidelity remains to be evaluated.

\begin{table}[t]\centering
\caption{Turn-taking statistics based on voice activity. Overlap and Switches/min are N/A for MOSS-TTSD (monaural output).}
\label{tab:turntaking}
\textbf{Test chunk Eval}\\[2pt]
\begin{tabular}{lrrr}
\toprule
System & Overlap & Silence & Switches/min \\
\midrule
GT  & 0.076 & 0.314 & 35.4 \\
KABURI (GT)   & 0.074 & 0.324 & 33.8 \\
KABURI (stat) & 0.105 & 0.199 & 35.6 \\
KABURI (pred) & 0.096 & 0.267 & 38.6 \\
Irodori (GT)   & 0.051 & 0.421 & 17.8 \\
Irodori (stat) & 0.013 & 0.402 & 11.6 \\
\bottomrule
\end{tabular}

\vspace{8pt}
\textbf{Text-in Eval}\\[2pt]
\begin{tabular}{lrrr}
\toprule
System & Overlap & Silence & Switches/min \\
\midrule
KABURI (stat) & 0.120 & 0.155 & 40.4 \\
KABURI (pred) & 0.161 & 0.166 & 43.8 \\
Irodori (stat) & 0.020 & 0.287 & 17.4 \\
MOSS-TTSD& N/A & 0.283 & N/A \\
\bottomrule
\end{tabular}
\vspace*{-5pt}
\end{table}
\section{Summary and Future Work}
\label{sec:summary}

We proposed KABURI-TTS, which takes a per-speaker phoneme raster as input and
generates two-party, one-speaker-per-channel dialogue speech conditioned on
voice activity, enabling controllable, human-like overlap. A human evaluation
and a voice-activity analysis confirmed higher naturalness than strong baselines
and more frequent overlap and turn-taking.

For future work, we aim to improve accuracy by learning a more human-like
placement. We also plan to compare KABURI-TTS with other speech synthesizers,
to explore backbones other than Irodori, and to support arbitrary durations.
Finally, we plan to use KABURI-TTS for data augmentation for full-duplex spoken
dialogue models. We also aim to adapt it to other languages, 
such as English and Chinese.

\section*{Acknowledgment}
This work was supported by the ``R\&D Hub Aimed at Ensuring Transparency and Reliability of Generative AI Models'' project of the Ministry of Education, Culture, Sports, Science and Technology.

\renewcommand*{\bibfont}{\fontsize{10}{11}\selectfont}
\printbibliography

\end{document}